\documentclass[preprint,showpacs,nofootinbib,preprintnumbers,amsmath,amssymb]{revtex4}
\usepackage{graphicx}

\begin{document}

\title{Transversity from two pion interference fragmentation}

\author{Jun She}
\author{Yang Huang}
\affiliation{School of Physics, Peking University, Beijing 100871,
China}

\author{Vincenzo Barone}
\affiliation{Di.S.T.A., Universit\`a del Piemonte Orientale ``A.
Avogadro'',  and INFN, Gruppo Collegato di Alessandria, 15100
Alessandria, Italy}

\author{Bo-Qiang Ma}
\email{mabq@phy.pku.edu.cn} \affiliation{School of Physics and State
Key Laboratory of Nuclear Physics and Technology, Peking University,
Beijing 100871, China}

\begin{abstract}
We present calculation on the azimuthal spin asymmetries for pion
pair production in semi-inclusive deep inelastic scattering (SIDIS)
process at both HERMES and COMPASS kinematics, with transversely
polarized proton, deuteron and neutron targets. We calculate the
asymmetry by adopting a set of parametrization of the interference
fragmentation functions and two different models for the
transversity. We find that the result for the proton target is
insensitive to the approaches of the transversity but more helpful
to understand the interference fragmentation functions. However, for
the neutron target, which can be obtained through using deuteron and
{$^3$He} targets, we find different predictions for different
approaches to the transversity. Thus probing the two pion
interference fragmentation from the neutron can provide us more
interesting information on the transversity.
\end{abstract}

 \maketitle

\section{Introduction}
At leading twist, the internal structure of the nucleon can be
described by three fundamental distribution functions. They are the
unpolarized, the longitudinal and the transversity distribution
functions. The former two have been well known, but the last one --
transversity~\cite{bdr}, is less known both theoretically and
experimentally. The difficulty lies in its chiral-odd property,
which makes it inaccessible in inclusive deep inelastic scattering
(DIS) process. However, transversity can manifest itself through
Collins mechanism~\cite{Collins1993} in single hadron production
where the chiral-odd distribution function (DF) couples with an also
chiral-odd fragmentation function (FF), the so called Collins
function. By observing the single spin asymmetry (SSA) phenomena, we
can extract the information on the transversity and the Collins
function. HERMES collaboration~\cite{hermes_single} and COMPASS
collaboration~\cite{compass_single} have already published their
data, reporting their observation of the non-zero SSA. Some
work~\cite{Anselmino} has been done to extract the transversity and
Collins function from the data. In future, JLab (Jefferson
Laboratory) has also planned to measure the transversity through the
same process but with the $^3$He target~\cite{JLab}. We expect
further exploration to give more information.

However, difficulty still exists for reliable measurements of
transversity. Since the transversity and the Collins function always
appear together in the single hadron production case, they in fact
cannot be directly measured independently. An alternative way to
measure transversity is to detect two unpolarized leading hadrons in
the final state, which was suggested first by Collins, Ladinsky,
Heppelmann~\cite{Collins_et} and then by Jaffe, Jin and
Tang~\cite{Jaffe}. In two hadron leptoproduction process, the
transversity gets factorized at leading twist through a new
chiral-odd FF, usually called the interference FF. The new
introduced FF is still unknown yet, but it can be cleanly measured
in $e^+e^-$ annihilation at Belle. Until now,
HERMES~\cite{hermes_double} and COMPASS~\cite{compass_double} have
already published their preliminary data on two hadron production
process with unpolarized beam and transversely polarized proton or
deuteron target, which made a first step in understanding the new
FF. On the theoretical side, some models have been put forward to
calculate the interference FF for $\pi^+\pi^-$ pair
production~\cite{Jaffe,Radici2002,Bacchetta2006}. In
Ref.~\cite{Bacchetta2006}, Bacchetta and Radici also gave their
prediction at HERMES and COMPASS kinematics, using different
parametrizations of transversity. In this paper, we will also give
predictions on SSA both at HERMES and COMPASS kinematics and with
various targets. For the interference FFs, we will adopt the
parametrization provided by Ref.~\cite{Bacchetta2006}, while for the
transversity, we will use two different models, the SU(6)
quark-diquark model and the pQCD based counting rule analysis.

\section{Cross sections and the asymmetry}

The asymmetry measured by the experiment is defined as:
\begin{eqnarray}
A_{UT}(\phi_{R},\phi_S,\theta) & = & \frac{1}{|S_T|}
\frac{N^{\uparrow}(\phi_{R},\phi_S,\theta)/N^{\uparrow}_\mathrm{DIS}
-N^{\downarrow}(\phi_{R},\phi_S,\theta)/N^{\downarrow}_\mathrm{DIS}}
{N^{\uparrow}(\phi_{R},\phi_S,\theta)/N^{\uparrow}_\mathrm{DIS}
 + N^{\downarrow}(\phi_{R},\phi_S,\theta)/N^{\downarrow}_\mathrm{DIS}} \nonumber \\
 & = &  \frac{\sigma_{UT}}{\sigma_{UU}}\mathrm{,}\label{eq:asymmetry}
\end{eqnarray}
where $UT$ refers to unpolarized beam and transversely polarized
target. The asymmetry is evaluated as a function of the angles
$\phi_{R}$, $\phi_S$ and $\theta$. $\phi_R$ denotes the azimuthal
angle of the detected two hadron plane with respect to the lepton
plane, and $\phi_S$ denotes the azimuthal angle of the polarization
vector $\vec{S}_T$ with respect to the lepton plane. $\theta$ is the
polar angle of the first hadron in the hadron pair's center-of-mass
frame with respect to the direction of the summed hadron momentum in
the lab frame\footnote{The angle definitions here are consistent
with the ``Trento Conventions''~\cite{Trento conventions}.}.

Consider the process $e \, \vec{N}\, {\longrightarrow} \, e' h_1 \,
h_2 \,X$, where the hadrons $h_1$ and $h_2$  are produced hadrons in
the current fragmentation region. An electron with momentum $l$
scatters off a proton target with mass $M$ and momentum $P$, via the
exchange of a virtual photon with momentum transfer $q=l-l^\prime$.
Inside the proton, a quark with initial momentum $p$ changes to a
state with momentum $k=p+q$ after the photon hit it. We define the
light-cone variable $x=p^+/P^+$, which represents the fraction of
target momentum carried by the quark. The detected two hadrons have
momenta $P_1$ and $P_2$, masses $M_1$ and $M_2$, and total invariant
mass $M_h^2 = (P_1 + P_2)^2$. We introduce the vectors $P_h=P_1+P_2$
and $R=(P_1-P_2)/2$, i.e., the total and relative momenta of the
hadron pair, respectively. We have
\begin{eqnarray}
|\vec{R}|=\frac{1}{2}\sqrt{M_h^2-2(M_1^2+M_2^2)}=\frac{1}{2}\sqrt{M_h^2-4m_\pi
^2},
\end{eqnarray}
if only $\pi^+\pi^-$ pairs are considered now. Similar to $x$, we
define $z=P_h^-/k^-$, which represents the fraction of fragmenting
quark momentum carried by the produced hadrons. We will also
introduce a light-cone fraction $\zeta={2R^-}/{P_h^-}$, which
describes how the total momentum of the pair is split into the two
hadrons,
\begin{eqnarray}
\label{zeta}
\zeta=\frac{2R^-}{P_h^-}=-\frac{2|\vec{R}|}{M_h}\cos\theta.
\end{eqnarray}

With the definitions above, the  cross section up to leading twist
can be expressed as:~\cite{Bacchetta2004}\footnote{Also see this
article for sub-leading twist expression.}
\begin{eqnarray}
\frac{d^7\sigma_{UU}}{d\zeta \,dM_h^2 \,d\phi_R \,dz \,dx \,dy
\,d\phi_S}~=&& \frac{\alpha^2}{2\pi Q^2 y}\,\sum_a e_a^2
A(y)\,f^a(x)\, D_1^a(z,\zeta, M_h^2),\\
\frac{d^7\sigma_{UT}}{d\zeta \,dM_h^2 \,d\phi_R \,dz \,dx \,dy
\,d\phi_S}~=&& -\frac{\alpha^2}{2\pi Q^2
y}|\vec{S}_T|\,\sum_a\,e_a^2 B(y)
\sin(\phi_R+\phi_S)\sin\theta\nonumber\\
&&\times\frac{|R|}{M_h} \delta f^a(x) H_1^{<\kern -0.3
em{\scriptscriptstyle )}\,a}(z,\zeta,M_h^2),
\end{eqnarray}
with $A(y)=1-y+y^2/2$ and $B(y)=1-y$. Here, $f(x)$ and $\delta f(x)$
denote the unpolarized and the transversity distribution functions
respectively. $D_1(z,\zeta,M_h^2)$ and $H_1^{<\kern -0.3
em{\scriptscriptstyle )}}(z,\zeta,M_h^2)$ are the new introduced
interference FFs, describing a quark fragmenting to a pair of
hadrons, for example, $\pi^+\pi^-$ pairs. After integration of
$\phi_R$, $\phi_S$ and $\zeta$, we define the weighted asymmetry:
\begin{eqnarray}
A_{UT}^{\sin(\phi_R+\phi_S)\sin\theta}(y,x,z,M_h^2)&=&\frac{2}{|\vec{S}_T|}\frac{\int
d\phi_S d\phi_R d\zeta \sin(\phi_R+\phi_S)/\sin\theta
d^7\sigma_{UT}}{\int d\phi_s d\phi_R
d\zeta d^7\sigma_{UU}}\nonumber\\
&=&-\frac{\frac{B(y)}{xy^2}\sum_a e_a^2 \delta f^a(x)\int d\zeta
\frac{|\vec{R}|}{M_h} H_1^{<\kern -0.3
em{\scriptscriptstyle)}\,a}(z,\zeta,M_h^2)}{\frac{A(y)}{xy^2}\sum_a
e_a^2 f^a(x) \int d\zeta D_1^a(z,\zeta,M_h^2)}.
\end{eqnarray}
More details on the interference FFs will be given in the next
section.

\section{Parametrization of distribution and fragmentation functions}
\subsection{Distribution Functions}
In this paper we will adopt two models: the SU(6) quark-diquark
model~\cite{Feynman,Close,Ma1996} and the pQCD based counting rule
analysis~\cite{Farrar,Blackenbecler,Brodsky,Gluck,Hirai} to get the
transversity distributions. Both two models have given pretty good
descriptions on the longitudinal polarized parton distribution
functions~\cite{Xun C}. A recent work~\cite{ourrecent} showed that
the prediction based on the two models for transversity was also
compatible with the current experiment data. We can say that both
models reflect the main feature of the nucleon structure in the
mediate $x$ region. But two models behave differently when
$x\rightarrow{1}$: the SU(6) quark-spectator-diquark
model~\cite{Ma1996} predicts $\delta{d(x)}/d(x)\rightarrow{-1/3}$,
while the pQCD based counting rule analysis~\cite{Brodsky} predicts
$\delta{d(x)}/d(x)\rightarrow {1}$. In a recent
literature~\cite{Anselmino}, Anselmino {\it et al.} extracted the
transversity distribution for $u$ and $d$ quarks from the now
available data, and showed some evidence that $\delta u(x)$ and
$\delta d(x)$ to be opposite in sign, with $|\delta d(x)|$ smaller
than $|\delta u(x)|$. This seems to be coincidence with the SU(6)
quark-diquark model qualitatively, but it clearly shows that $\delta
d(x)/d(x)\rightarrow 0$ when $x\rightarrow 1$, which is coincidence
with neither model we used in this paper at large $x$ region. The
correctness of different parametrization is still unclear, and need
to be checked by more experiments.

For the SU(6) quark-diquark model, we will adopt one set of the
unpolarized quarik distribution parametrization as a input, and then
use theoretical relations to connect the quark transversity
distributions with the unpolarized
distributions~\cite{Ma1996,Ma_plb1998}:
\begin{eqnarray}
&&\delta
u_v(x)=[u_v(x)-\frac{1}{2}d_v(x)]\hat{W}_S(x)-\frac{1}{6}d_v(x)\hat{W}_V(x),
\nonumber\\
&&\delta{d}_v(x)=-\frac{1}{3}d_v(x)\hat{W}_V(x),
\end{eqnarray}
$\hat{W}_S(x)$ and $\hat{W}_V(x)$ are the Melosh-Wigner rotation
factors~\cite{Ma_plb1998,Ma1998, Schmidt and Soffer1998} for
spectator scalar and vector diquarks, which come from the
relativistic effect of quark transversal motions~\cite{Ma1991}. This
model predicts $d_v(x)/u_v(x) \rightarrow 0$ when $x \rightarrow 1$,
which is compatible with the available experiment data.

For the pQCD based analysis, we adopt the parametrization
\begin{eqnarray}
u_v^{pQCD(x)}=u_v^{para}(x),~~~~~~d_v^{pQCD}(x)=\frac{d_v^{th}(x)}{u_v^{th}(x)}u_v^{para}(x),\nonumber\\
\delta u_v^{pQCD}(x)=\frac{\delta
u_v^{th}(x)}{u_v^{th}(x)}u_v^{para}(x),~~~~~~\delta
d_v^{pQCD}(x)=\frac{\delta d_v^{th}(x)}{u_v^{th}(x)}u_v^{para}(x),
\end{eqnarray}
where the superscripts ``th'' means the theoretical calculation in
the pQCD analysis~\cite{Ma037501,Ma014017}, and ``para''means the
input from parametrization. The pure theoretical calculation in this
model predicts that $d_v(x)/u_v(x) \rightarrow 1/5$ when $x
\rightarrow 1$. So we use a factor $u^{para}_{v}(x)/u^{th}_{v}$ to
adjust each pure theoretically calculated quantity to a more
realistic pQCD model quantity.

In this paper, we will use the CTEQ6L~\cite{CTEQ}
parametrization\footnote{This parametrization gives that
$d_v(x)/u_v(x) \rightarrow 0$ when $x \rightarrow 1$, which is
coincidence with the current data.} as the input for both models to
get the unpolarized parton distribution functions. Detailed
constructions of the quark distributions can be found in
Ref.~\cite{Ma037501,Ma014017,Ma114009}.

\subsection{Interference Fragmentation Functions}
The so called interference FFs $D_1(z,\zeta,M_h^2)$ and $H^{<\kern
-0.3 em{\scriptscriptstyle )}}_{1}(z,\zeta,M_h^2)$ describe a quark
splitting into a pair of unpolarized hadrons inside the same jet.
Different models~\cite{Jaffe,Radici2002,Bacchetta2006} have given
their calculated results on the interference FFs. In this paper, we
will follow the parametrization given by Ref.~\cite{Bacchetta2006},
where they used the spectator model to get the result.

From Eq.~\ref{zeta}, we find that the dependence on $\zeta$ can be
expressed on $\cos\theta$. Expanding the hadron pair system in
relative partial waves, we get:~\cite{Bacchetta2003}
\begin{eqnarray}
D_1^a(z,\cos\theta,M_h^2)&\approx&D_{1,\,oo}^a(z,M_h^2)+D_{1,\,ol}^a(z,M_h^2)
\cos\theta+D_{1,\,ll}^q(z,M_h^2)\frac{3\cos^2\theta-1}{4},\\
H^{<\kern -0.3 em{\scriptscriptstyle )}\,a}_{1}(z,
\cos\theta,M^2_h)&\approx&H^{<\kern -0.3 em{\scriptscriptstyle
)}\,a}_{1,\,ot}(z,M_h^2)+H^{<\kern -0.3 em{\scriptscriptstyle
)}\,a}_{1,\,lt}(z,M_h^2)\cos\theta.
\end{eqnarray}
Integrating over $\zeta$, i.e., the $\cos\theta$, we can easily find
that only $D_{1,oo}^a$ and $H^{<\kern -0.3 em{\scriptscriptstyle
)}\,a}_{1,ot}$ contribute to the final result. The factor
$|\vec{R}|/M_h$ appearing due to the Jacobian can be absorbed in the
definition of integrated interference FFs. The explicit expressions
for $D_{1,\,oo}^a$ and $H^{<\kern -0.3 em{\scriptscriptstyle
)}\,a}_{1,\,ot}$ can be found in Ref.~\cite{Bacchetta2006}.

\section{Numerical Calculations}
We present the final formula for calculating the asymmetry:
\begin{eqnarray}
\label{asy} A_{UT}^{\sin(\phi_R+\phi_S)\sin\theta}(y,x,z,M_h^2)=
-\frac{\frac{1-y}{xy^2}}{\frac{1-y+y^2/2}{xy^2}}
\frac{|\vec{R}|}{M_h} \frac{\sum_a e_a^2 \delta
f^a(x)H_{1,\,ot}^{<\kern -0.3
em{\scriptscriptstyle)}\,a}(z,M_h^2)}{\sum_a e_a^2
f^a(x)D_{1,\,oo}^a(z,M_h^2)}.
\end{eqnarray}
By integrating through various ways on the numerator and
denominator, we can get the asymmetry depending on different
kinematical variables. In this paper, the dependencies on $M_h$, $x$
and $z$ are calculated.

For each target, we will perform the numerical calculations under
both HERMES and COMPASS experiment cuts. In the HERMES experiment,
the kinematical cuts are:
\begin{eqnarray}
Q^2 >1 ~\textmd{GeV}^2,\ W > 2~\textmd{GeV},\ 0.1 < y < 0.85,\ 0.2 <
z < 0.7,\ 0.5<M_h<1~\textmd{GeV}.
\end{eqnarray}
For the $Q^2$ and $W$ used in the integration over $y$ and $x$, we
use the relations
\begin{eqnarray}
Q^2=sxy,~~W^2=sy(1-x)+M^2,
\end{eqnarray}
with $s=2ME=51.8 \textmd{GeV}^2$ in the HERMES experiments.

For COMPASS, the kinematics are:
\begin{eqnarray}
&&s=300~\textmd{GeV}^2,~Q>1.0~\textmd{GeV},~W>5.0~\textmd{GeV},\nonumber\\
&&0.1<y<0.9,~0.1<z<0.9,~0.3<M_h<2.5~\textmd{GeV}.
\end{eqnarray}
We notice first that the beam energy is extremely high ($\mu^+$ beam
with 160GeV) that the COMPASS experiment can detect very small $x$
region, so for convenience, we will adopt the logarithm coordinate.
Second, COMPASS can reach a higher $M_h$ than HERMES, but the model
for the interference FFs does not consider the contributions from
resonances with higher masses. We argue that the parametrization
should be modified for higher invariant mass of the pair system, so
here we only present the calculation up to the HERMES cut for $M_h$.

\subsection{Proton target}
\begin{figure} \center
\includegraphics[scale=1.5]{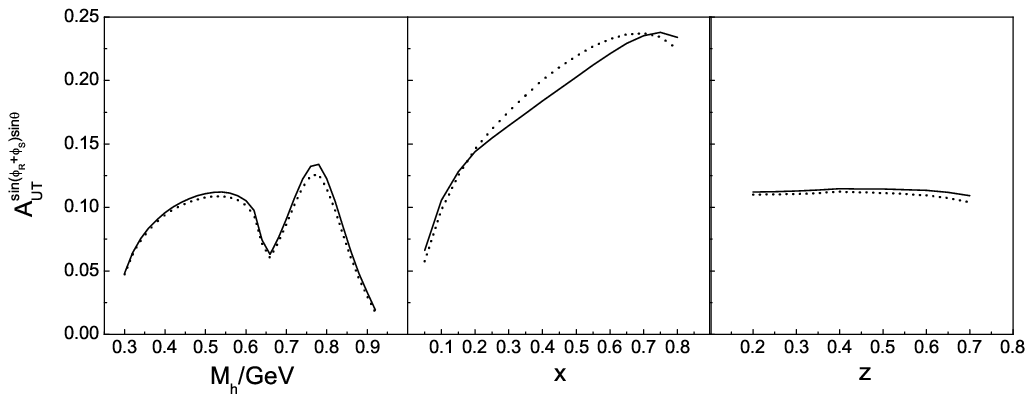}
\caption{$A_{UT}^{\sin(\phi_R+\phi_S)\sin\theta}$ at HERMES
kinematics for a transversely polarized proton target as a function
of $M_h$, $x$ and $z$ respectively. The solid lines and dotted lines
correspond to the results evaluated from SU(6) quark-diquark model
and pQCD based counting rules, respectively.} \label{phermes}
\includegraphics[scale=1.5]{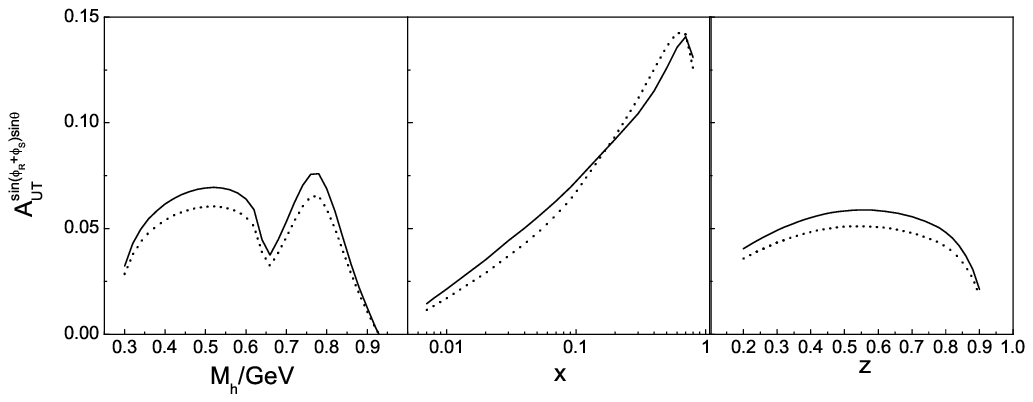}
\caption{The same as Fig.~\ref{phermes}, but at COMPASS kinematics}
\label{pcompass}
\end{figure}

The numerical result for proton target are shown in
Fig.~\ref{phermes} and Fig.~\ref{pcompass}.

From Fig.~\ref{phermes} and Fig.~\ref{pcompass}, we can see that
different models for the transversity give almost the same
predictions on the asymmetry for the proton target. This is because
the proton target is dominated by $u$ quarks, and the two models
give similar predictions on $u$ quark distributions~\cite{Ma034010}.
Besides this, the contribution from $u$ quarks should be magnified
by 4 times due to the charge. So we conclude that the result is
insensitive to the models of transversity for proton target, thus
from this experiment, we cannot distinguish the two models. However,
in the mediate $x$ region, this gives us a chance to measure the
unknown interference FFs, which is helpful to explore the new
domain. Now, different models give different predictions on the
interference FFs. According to Ref.~\cite{Jaffe}, the FF was
anticipated to change sign around $\rho$ mass, while in
Ref.~\cite{Bacchetta2006,Radici2002}, they predicted a peak at
$\rho$ mass. Even between the results in Ref.~\cite{Radici2002} and
Ref.~\cite{Bacchetta2006}, there are also slightly differences. If
the conclusion that the asymmetry is insensitive to different
approaches of the transversity holds, we expect the experiments to
publish more data on proton target to clarify the details on the
interference FFs.

\subsection{Deuteron target}
\begin{figure}
\center
\includegraphics[scale=1.5]{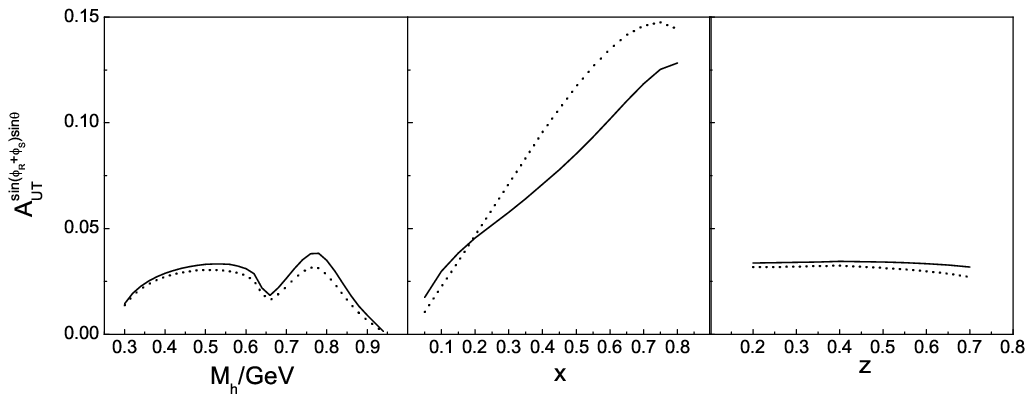}
\caption{The same as Fig.~\ref{phermes}, but for deuteron target.}
\label{dhermes}
\includegraphics[scale=1.5]{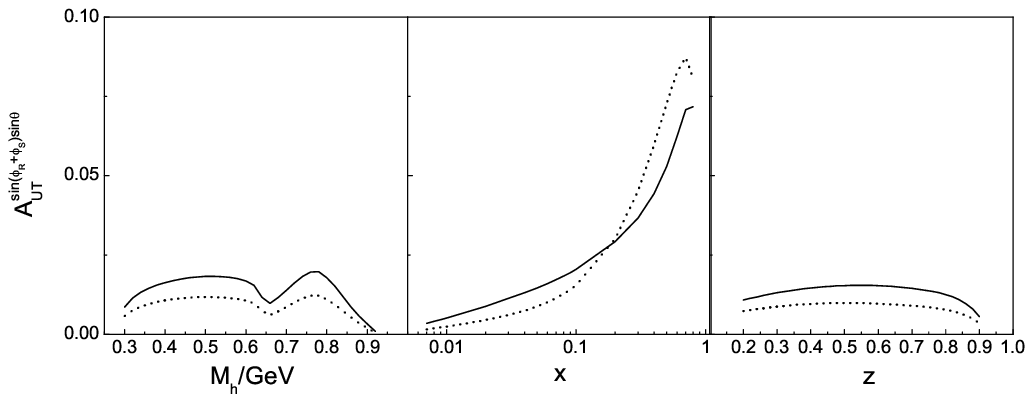}
\caption{Similar to Fig.~\ref{pcompass}, but for deuteron target.}
\label{dcompass}
\end{figure}

The result for deuteron target is shown in Fig.~\ref{dhermes} and
Fig.~\ref{dcompass}.
Inside the deuteron, the $u$ and $d$ quarks have the same
distribution, and because of the charge, $u$ quarks still dominant
here, and the asymmetry is still not so sensitive to different
models of the transversity. So the deuteron target can also be used
to measure the interference FFs. Combining the experiment data from
the proton and deuteron targets, we can get abundant information not
only on the interference FFs but also the transversity
distributions, especially for $u$ quarks.

\subsection{Neutron target}
\begin{figure}
\center
\includegraphics[scale=1.5]{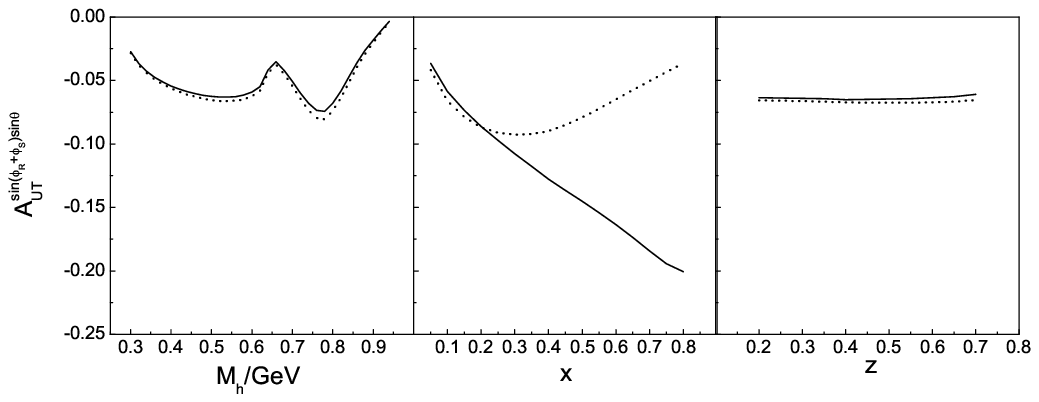}
\caption{The same as Fig.~\ref{phermes}, but the neutron target is
assumed here.} \label{nhermes}
\includegraphics[scale=1.5]{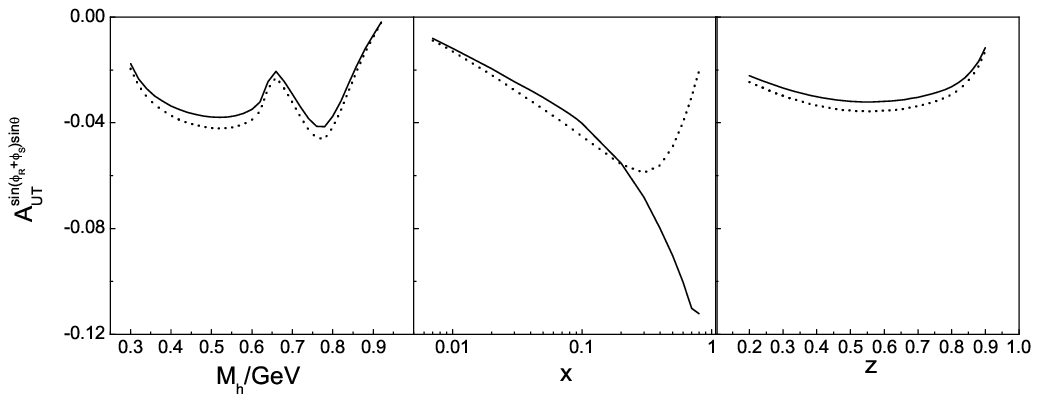}
\caption{The same as Fig.~\ref{pcompass}, but the neutron target is
assumed here.} \label{ncompass}
\end{figure}

Although from the data on deuteron target, we can get a first glance
at the transversity distribution for $d$ quarks, we suggest a
directly measurement using the neutron target. Fig.~\ref{nhermes}
and Fig.~\ref{ncompass} show the result on the neutron target.
Due to the fact that the information on $z$-dependence and
$M_h$-dependence is only contained in the interference FFs, two
models should give nearly the same prediction on the $z$ and $M_h$
dependence of the asymmetry, which is similar to the proton and
deuteron targets. But for $x$-dependence, because $d$ quarks
dominate in the neutron target, and two models give quite different
predictions on the $d$ quark distributions~\cite{Ma034010}, the two
models might exhibit their differences here, even if the
contribution from $d$ quarks is suppressed by a factor of 4
originated from the square of the electric charge compared with $u$
quarks. As the figures show, different models for the transversity
predict differently on the asymmetry when $x$ increases, thus this
is helpful for us to distinguish the two models. However, we should
notice that this effect is apparent only in the large $x$ region.
Both HERMES and COMPASS did the experiment in the relative low $x$
region, and this difference is not so obvious there. So we expect
further experiments will reach higher $x$ region to help us
distinguish the models. Another problem is that it is difficult to
acquire free neutron target, so $^3$He target is suggested, which
can be considered as an effectively free neutron, because two
protons inside the nucleus form a spin singlet. JLab has planned
measurements on the $^3$He target, so we look forward to the result
from JLab.

Careful analysis with data from both proton and deuteron targets may
also provide an extraction of neutron result. This can be done by
combining both HERMES and COMPASS experiments, or COMPASS perform
precision measurements on both proton and deuteron targets
respectively.

Unlike the case in single pion production where the the predictions
on the neutron are insensitive to different models~\cite{ourrecent},
the double pion production are ideal to distinguish between
different model predictions. The first reason is that there is no
dilation caused by unfavored fragmentation functions as in the
single pion case, so that the contribution from the $d$-distribution
of the nucleon (in fact it is $u$-distribution in the neutron) can
manifest itself more clearly in double pion interference
fragmentation. Another important reason is that the two pion
interference fragmentation function causes the $d$-quark
contribution to have an opposite sign compared to that of the
$u$-quark contribution in the single spin asymmetry
formula~\cite{Bacchetta2006}, so that the calculated single spin
asymmetries are always negative in both the two models for the
neutron case. More explicitly, we can predict the large $x$-behavior
\begin{eqnarray}
&&A_n=-\frac{1}{9}\times \frac{21}{19} A_p , ~~~~~~   \mbox{for pQCD inspired model}; \\
&&A_n=-A_p,  ~~~~~~~~~~~~~~~~ \mbox{for quark-diquark model},
\end{eqnarray}
at $x \to 1$ for the single spin asymmetry. This provides a strong
motivation to do experiments on extracting the neutron result of
single spin asymmetry in double hadron production.

\section{Summary}
Transversity distribution is the less known piece in understanding
the spin structure of the nucleon due to its chiral-odd nature.
Single spin asymmetry of single hadron production in semi-inclusive
deep inelastic scattering (SIDIS) provide a way accessing the
transversity, in which the transversity distribution couples with an
also chiral-odd fragmentation function (FF), the collins function.
Another interesting way to measure the transversity is through
observing single spin asymmetry (SSA) in double hadron production,
where transversity gets factorized with the so called interference
FF. One advantage for this method is that the interference FF can be
measured separately in the $e^+e^-$ annihilation process, so that we
can get a clean result on transversity. HERMES has already finished
the experiment and will publish their data in near future. COMPASS
has also published their preliminary data and is still accumulating
data. In this paper, we present numerical calculation for the
proton, deuteron and neutron target respectively at the HERMES and
COMPASS kinematic region, using two models for transversity and a
set of parametrization of interference FFs provided by
Ref.~\cite{Bacchetta2006}. We found that two models, the SU(6)
quark-diquark model and the pQCD based counting rule analysis give
quite similar prediction at HERMES and COMPASS kinematics for the
proton and deuteron target, i.e., the result is insensitive to
different approaches of the transversity. Thus we argue that the
HERMES and COMPASS experiment can provide rich information on the
interference FFs.
For the neutron target, we found that the two models give different
predictions at large $x$ region, which is helpful to distinguish
them. So we suggest doing experiments with $^3$He target (an
effective neutron target) at large $x$ region to give more
information, especially that on the transversity of $d$ quarks.
Maybe JLab will bring us exciting results. Careful analysis of data
from both proton and deuteron targets by HERMES and COMPASS might be
also useful to extract the neutron result, for the sake to confront
different theoretical predictions on transversity.

\section{acknowledgement}
We are grateful to Xiaorui Lu and Gunar Schnell for useful
discussion. This work is partially supported by National Natural
Science Foundation of China (Nos.~10421503, 10575003, 10528510), by
the Key Grant Project of Chinese Ministry of Education (No.~305001),
by the Research Fund for the Doctoral Program of Higher Education
(China), and by the Italian Ministry of University and Research
(PRIN 2007).

\end{document}